# Visible Integrated Photonics with Tantalum Pentoxide

Sergei Nedić[1,2], Angus Gale[1*], Hugo Quard[1,2], Nathan Coste[1,2], Anastasiia Zalogina[1,2], Igor Aharonovich[1,2,*]

*[1] School of Mathematical and Physical Sciences, Faculty of Science, University of Technology Sydney, Ultimo, New South Wales 2007, Australia*

*[2] ARC Centre of Excellence for Transformative Meta-Optical Systems (TMOS), University of Technology Sydney, Ultimo, New South Wales 2007, Australia*

*igor.aharonovich@uts.edu.au

*angus.gale@uts.edu.au

*Developing chip-scale photonic platforms for the visible spectrum is essential for translating next-generation optical technologies. Here, we demonstrate that tantalum pentoxide $Ta_2O_5$ on insulator is a highly promising material framework for these applications, supporting high-Q resonator architectures across the visible spectral range. Utilizing routine and scalable thin-film deposition and fabrication methodologies, we realize $Ta_2O_5$ waveguide and resonator layouts compatible with photonic integrated circuit (PIC) architectures. Through a comparison with numerical simulations, we characterize compact racetrack resonators, evaluating their performance in both the green and blue spectral regions. In addition, we fabricate large-scale ring resonators exhibiting record-high quality factors for $Ta_2O_5$ in the visible spectrum, exceeding $7.0 \times 10^4$. These results highlight the immense potential of this platform for visible photonic systems and establish a robust, straightforward baseline for future visible integrated optics.*

Photonic integrated circuits (PICs) are emerging as a powerful platform for next-generation information and communication technologies[1–4]. PICs feature micron-scale device footprints and offer the potential for large scale fabrication processes. In combination with their high component density, these attributes enable exceptional design flexibility, scalability, and cost efficiency across a broad range of applications[5–7]. Desirable characteristics for a PIC material platform include compatibility with established fabrication technologies and a sufficiently wide bandgap to minimize optical absorption across wavelengths. Additional important characteristics include low autofluorescence, optical nonlinearity, a high refractive index, and the ability to integrate with quantum emitters (QEs)[5,8–12].
Several photonic platforms have been explored for on-chip integration with these properties in mind. Silicon (Si) is widely used due to its mature fabrication technology[13], high refractive index, and ability to host quantum emitters[14]. More recently, silicon nitride ($Si_3N_4$) has attracted attention for its low propagation loss and strong optical nonlinearity[15–17], while aluminium nitride (AlN), with its extremely wide bandgap of ~6.015 eV, offers unique opportunities for ultraviolet photonics[11].
Among emerging material platforms, tantalum pentoxide ($Ta_2O_5$) has gained attention as a CMOS-compatible dielectric exhibiting exceptionally low propagation losses, reaching 3 ± 1 dB/m at 1550 nm[18–22]. With an optical bandgap of approximately 3.9 eV, $Ta_2O_5$ shows negligible absorption from the ultraviolet to the mid-infrared spectral range[23]. Combined with its intrinsically low autofluorescence, these properties make $Ta_2O_5$ an excellent candidate for low-noise integration of QEs at short wavelengths. Its low thermo-optic coefficient ($1.14 \times 10^{-6}$ RIU/K) is particularly advantageous for applications requiring minimal thermo-refractive noise, such as superconducting nanowire single-photon detectors, where $Ta_2O_5$ waveguides have been integrated with NbTiN nanowires operating at 1.3 K[24]. In addition, its nonlinear response has been exploited in optical parametric oscillators based on hybrid waveguide–fiber architectures, while its dielectric properties have enabled nano-electromechanical phase shifters[25].

In this work, we demonstrate the capabilities of the $Ta_2O_5$ integrated photonics platform for visible-wavelength applications through the fabrication of compact racetrack resonators compatible with scalable photonic integrated circuit (PIC) architectures. Fabricated devices are compared directly with numerical simulations to extract real-world performance metrics. We investigate devices across a range of bus-resonator spacing to identify their coupling regime and to extract the corresponding coupling and transmission coefficients in both the green and blue spectral regions. In addition, we design and realise large-scale ring resonators exhibiting record-high quality factors for $Ta_2O_5$ in the visible spectrum, exceeding $7.0 \times 10^4$, highlighting the potential of this platform for low-loss, high-performance visible photonic systems.

To demonstrate the capabilities of the $Ta_2O_5$ photonic integrated circuit (PIC) platform, we fabricate bus-coupled racetrack and ring resonators, utilizing directional grating couplers for efficient fiber in- and out-coupling. Figure 1(a) shows the device architecture of the racetrack resonators studied in this work. The geometry comprises two semicircular waveguides of radius $R$ connected by straight sections of length $L$, forming a closed resonant cavity. A bus waveguide is placed at a separation gap, $g$, to enable evanescent coupling and is terminated with grating couplers at both ends for optical access. This design allows controlled tuning of the coupling strength via $L$ and $g$. Devices are fabricated from a 200 nm-thick $Ta_2O_5$ film with 300 nm-wide waveguides unless otherwise specified. Figures 1(b–d) show scanning electron microscope (SEM) images of the device array, a representative resonator, and a grating coupler.

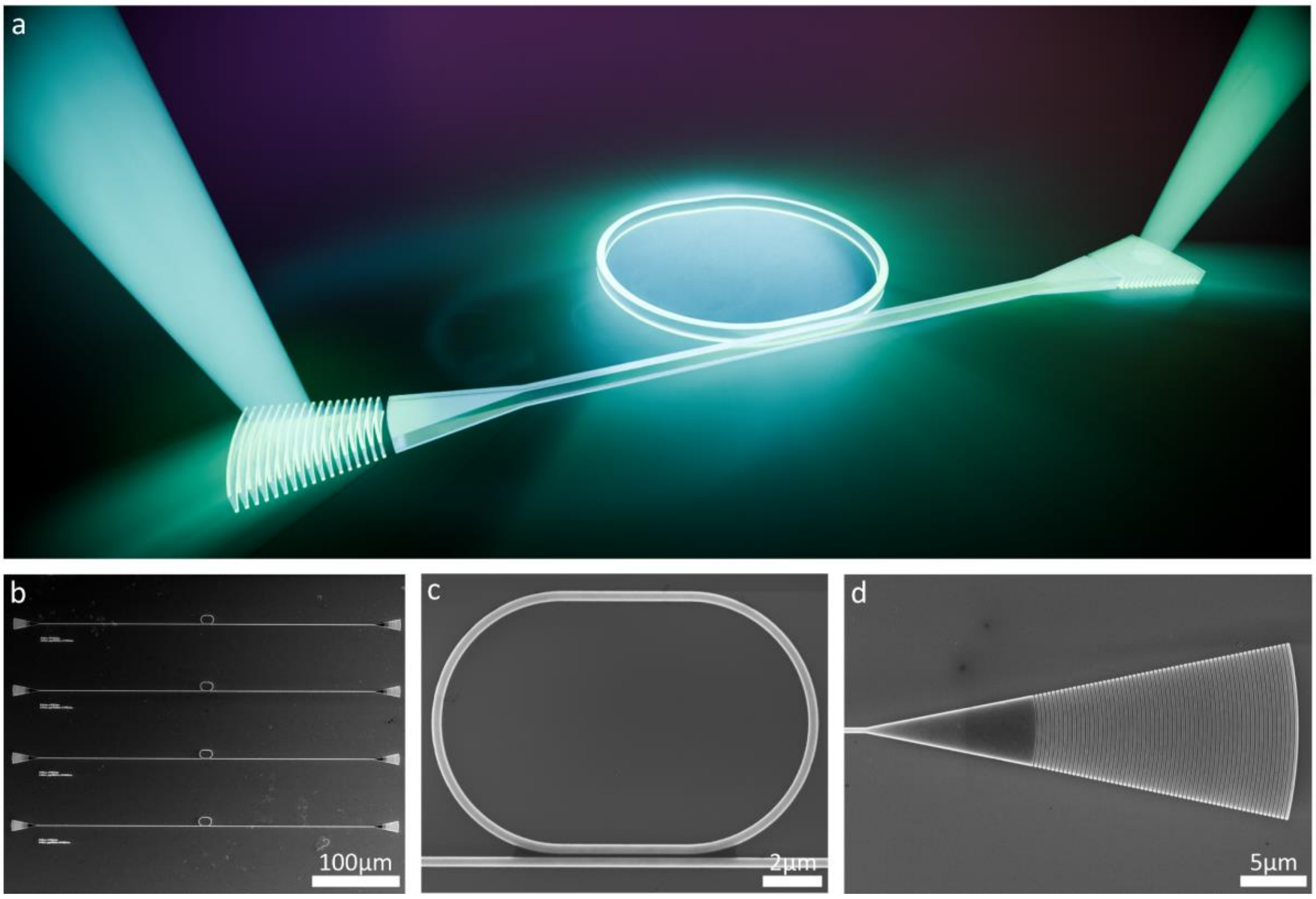

*Figure 1. (a) Three-dimensional schematic of the $Ta_2O_5$ racetrack resonator integrated with a bus waveguide and grating couplers for optical input and output. (b) Scanning electron microscope (SEM) image of an array of $Ta_2O_5$ racetrack resonators with a radius of 5 µm. (c) Enlarged SEM image of an individual racetrack resonator and (d) grating coupler.*

We performed 3D finite-difference time-domain (FDTD) simulations of a bus waveguide evanescently coupled to a racetrack resonator. The transmission spectrum was calculated by monitoring the optical power of the fundamental $TE_{00}$ mode at an output port located at the end of the bus waveguide. Resonant modes appear as dips in transmission, corresponding to efficient coupling of light into the resonator.

Figure 2(a) shows the simulated transmission spectrum of a racetrack resonator with $R$ = 5 µm, $L$ = 5 µm, and waveguide cross section of 300 nm width and 200 nm height. The gap, $g$ between the ring and the bus waveguide is 280 nm, and the waveguide shares the same width and thickness as the ring. Multiple resonances are observed between 558 and 565 nm, consistent with whispering-gallery modes supported by the resonator. An equivalent simulation performed in the blue spectral region is shown in Figure S3. The waveguide group index, $n_g$ can be determined using the cavity dispersion relation[26]:

$$n_g = \frac{\lambda_1 \lambda_2}{FSR \cdot P} \qquad (1)$$

where $\lambda_1$ and $\lambda_2$ are the centre wavelengths of two adjacent resonant modes, FSR is the free spectral range between the two modes ($\lambda_1$ - $\lambda_2$) and $P$ is the round-trip length of the racetrack resonator. An FSR of ~3.2 nm is seen from Figure 2(a), giving $n_g$ = 2.38. Notably for $g$ = 280 nm the resonator is significantly undercoupled with a normalised resonance transmission of ≈ 0.94. Extracted quality factor for the resonance at 560.1 nm yields a quality factor, $Q \approx 17300$, indicating moderate confinement and low intrinsic losses even for a resonator with small $R$. Figure 2(b) presents the transmission spectrum for the same racetrack structure with a reduced coupling gap, $g$ = 190 nm. As the system is now in an overcoupled regime, the normalised transmission on resonance drops significantly to ≈ 0.44 for the same resonant mode. Additionally, the simulated $Q$ is reduced to ≈ 13600, a consequence of increased coupling from the bus to resonator.

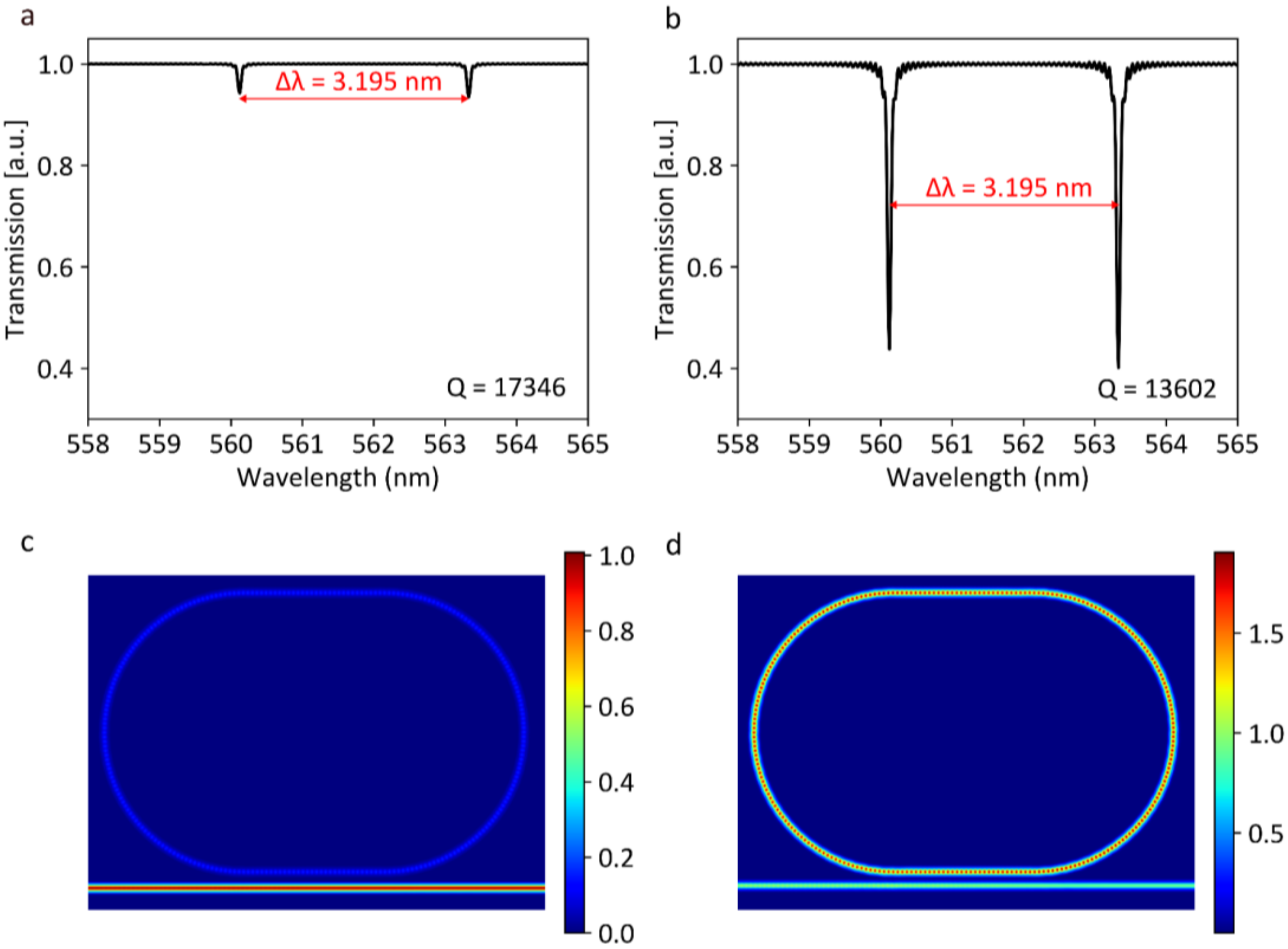

*Figure 2. (a,b) 3D FDTD-simulated transmission spectra of $Ta_2O_5$ racetrack resonators with radius R = 5 µm, coupling length L = 5 µm, and waveguide dimensions of 200 nm × 300 nm (height × width) on $SiO_2$. The coupling gaps are (a) 280 nm, corresponding to an undercoupled resonator with Q ≈ 17300, and (b) 190 nm, corresponding to an overcoupled resonator with Q ≈ 13600. Both devices exhibit a free spectral range (FSR) of 3.195 nm. (c,d) In-plane electric field intensity distributions at the 560.1 nm resonance wavelength for coupling gaps of (c) 280 nm and (d) 190 nm, showing the weaker and stronger coupling into the resonator, respectively.*

Figures 2(c,d) provide further insight into the coupling behaviour through the simulated in-plane electric field distributions at a resonance wavelength of 560.1 nm. For $g$ = 280 nm (Figure 2(c)), only weak field buildup is observed within the resonator, consistent with an undercoupled regime in which most of the optical power remains in the bus waveguide. Reducing the gap to $g$ = 190 nm (Figure 2(d)) increases the coupling strength and drives the device into the overcoupled regime, resulting in pronounced field enhancement within the resonator and reduced transmitted power at the output port.

These regimes can be understood in terms of the round-trip amplitude transmission factor, $a$, and the self-coupling coefficient, $r$. An undercoupled resonator is characterised by $a < r$, indicating that intrinsic cavity losses exceed the coupling rate. At critical coupling $a = r$, the power coupled into the resonator exactly balances the round-trip losses, producing complete destructive interference at the through port. In the overcoupled regime $a > r$, coupling dominates over intrinsic loss, leading to increased intracavity power and a recovery of the resonant transmission dip. By combining analytical expressions for the quality factor, $Q$, and the minimum transmission, $T_{\min}$, at resonant wavelength $\lambda$, the parameters $a$ and $r$ can be extracted simultaneously according to[26]:

$$T_{min} = \left(\frac{r-a}{1-ra}\right)^2 \quad (2)$$

$$Q = \frac{\pi \, n_g \, P \sqrt{ra}}{\lambda \, (1-ra)} \quad (3)$$

For the resonant mode at 560.1 nm in the undercoupled regime ($g$ = 280 nm), substituting simulated values of $T_{\min}$ = 0.94 and $Q$ = 17346 yields a round-trip amplitude transmission factor $a$ = 0.96 and self-coupling coefficient $r$ = 0.99. In contrast, for the overcoupled system ($g$ = 190 nm), with $T_{\min}$ = 0.44 and $Q$ = 13602, the parameters for this same resonance are determined to be $a$ = 0.96 and $r$ = 0.89. These extracted values validate the coupling regimes, whereby $a < r$ for undercoupling and $a > r$ for overcoupling.

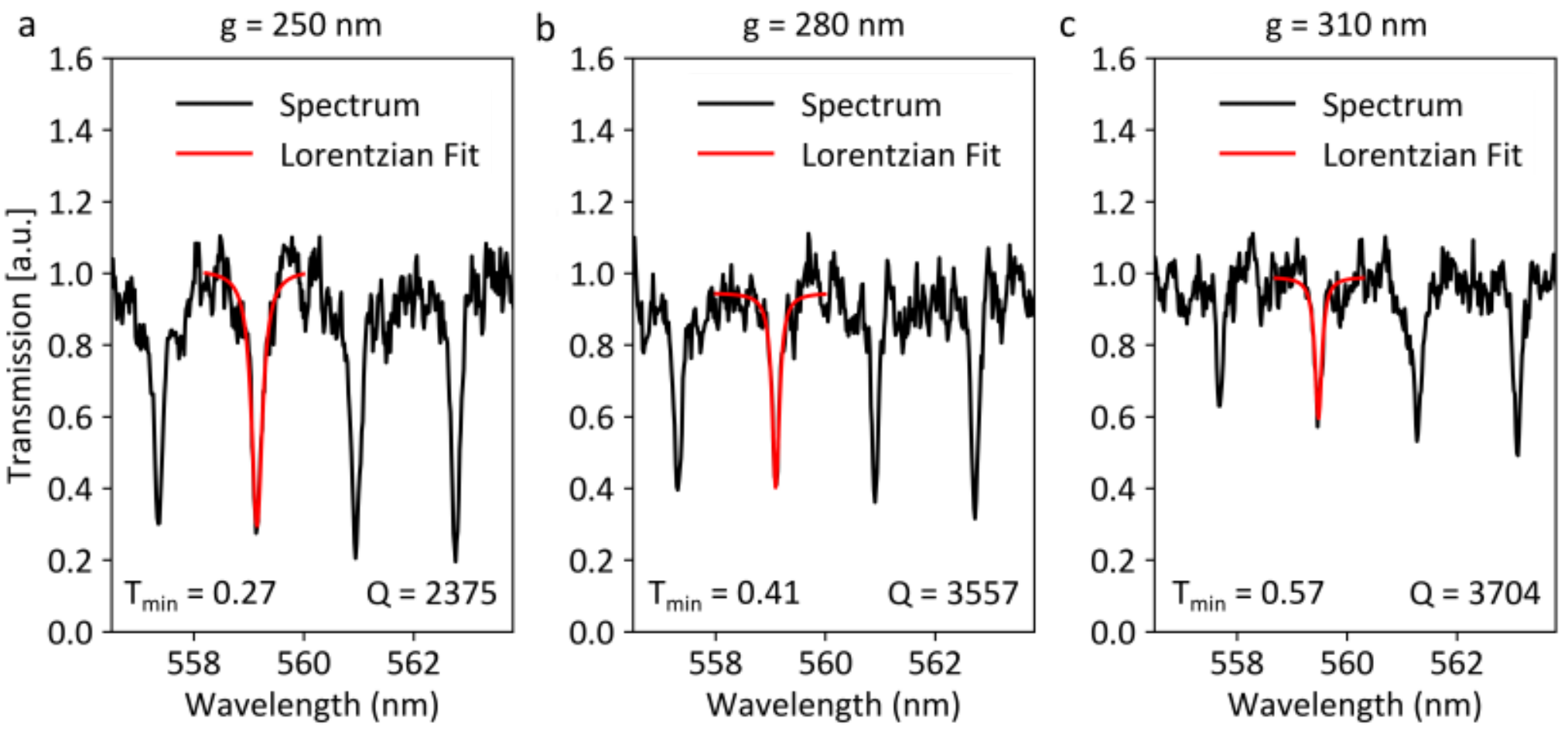

*Figure 3. (a–c) Transmission spectra of $Ta_2O_5$ racetrack resonators with R = 10 µm and L = 5 µm. The waveguide cross-section is 200 nm × 300 nm (height × width), with coupling gaps, g, of (a) 250 nm, (b) 280 nm, and (c) 310 nm. Periodic resonances corresponding to the cavity modes are observed.*

We now focus on the experimental measurements taken from a series of racetrack resonators ($R$ = 10 µm, $L$ = 5 µm) with varying only the coupling gap, $g$. Figure 3 shows background corrected transmission spectra (details in Figure S2) from 556 - 564 nm, for the same devices with $g$ = 250 nm, 280 nm and 310 nm. As expected, for increasing $g$, the value for $T_{min}$ is increased from 0.27 - 0.57. This is also accompanied by a modest increase in $Q$ from 2400 - 3700. An average FSR for all three rings is also measured at 1.80 nm.

Utilising the resonant cavity dispersion relation (Eqn. 1), $n_g$ is calculated as 2.40. By following the framework established using Equations 2 and 3, experimental values for both $a$ and $r$ can be extracted for all 3 devices. The resulting ($a$, $r$) values progress systematically as the coupling gap widens, (0.73, 0.91) for $g$ = 250 nm, (0.80, 0.95) for $g$ = 280 nm, and (0.79, 0.97) for $g$ = 310 nm. The increasing $r$ coefficient with increasing gap is a clear indication that all the resonators remain in the undercoupled regime. There is also a significant reduction in the round-trip amplitude transmission factor, $a$, compared to simulated results presented in Figure 2. For both $g$ = 280 nm and 310 nm $a \approx 0.80$ indicating that losses in the ring are noticeably increased. We note that for $g$ = 250 there is a further reduction in $a$ to 0.73. The narrower gap enhances the evanescent field overlap within the coupling region (indicated by reduced $r$ = 0.91), increasing the optical mode's interaction with the etched sidewall and inducing a small round-trip attenuation penalty. Ellipsometric measurements of the bulk deposited film point to negligible absorption (See Figure S1), and bending losses are reduced with the doubled racetrack radius, $R$. This then points to the impact of scattering losses which are especially prominent at green wavelengths. Unlike the simulated structures from Figure 2, real world devices suffer from fabrication inconsistencies such as sidewall roughness. To estimate the losses, we assume that scattering is the dominant loss mechanism. Following the cavity definitions in 26, the round-trip amplitude transmission factor, $a$, can be used to define the total fractional power loss per round-trip, $\alpha_s$, as:

$$\alpha_s = 1 - a^2 \qquad (4)$$

From the experimental transmission, $a \approx 0.80$ yields a total power attenuation of $\alpha_s$ = 0.36. This indicates that approximately 36% of the optical power is scattered or radiated out of the cavity mode on every round trip. The simulated result for a smaller $R$ = 5 um, gives $a$ = 0.96, which corresponds to an inherent round-trip power loss of just under 8%. Comparing our experimental results to this baseline clearly isolates the penalty imposed by real-world fabrication imperfections at these wavelengths. The drop in experimental $Q$ is a direct consequence of this power loss, showing that sidewall scattering, compounded with the compact geometry of these devices, sets a structural limit on performance.

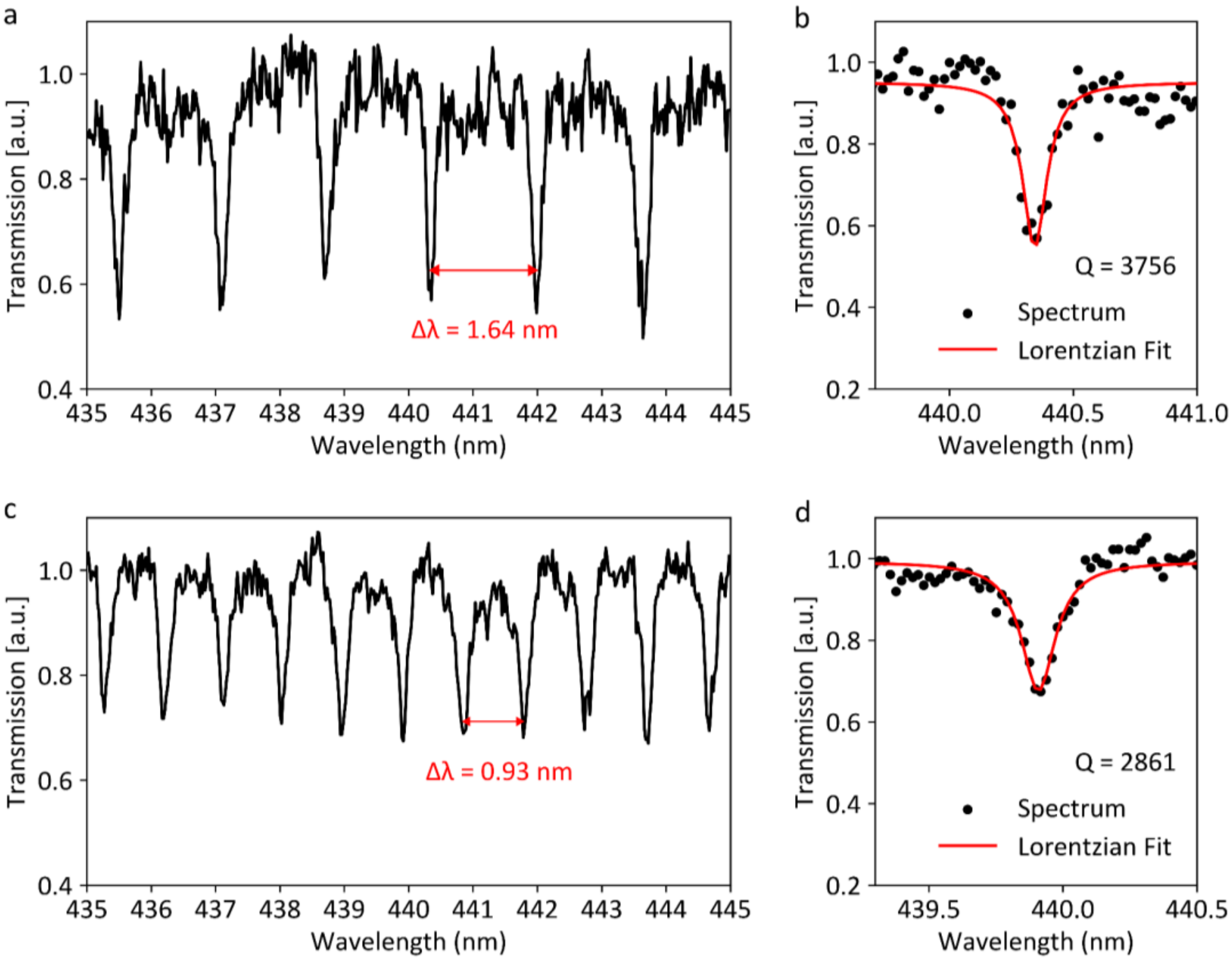

*Figure 4. (a,c) Transmission spectra of $Ta_2O_5$ racetrack resonators (L = 5 μm, g = 250 nm) with increasing round-trip length (P), achieved by increasing the racetrack radius from (a) R = 5 μm to (c) R = 10 μm. (b,d) Lorentzian fits to the resonant modes at (b) 440.3 nm and (d) 439.9 nm, yielding quality factors of Q = 3756 and Q = 2861, respectively.*

While there are a number of material platforms that have already been demonstrated for red visible wavelengths[17,27,28], shifting to blue wavelengths presents a number of challenges. These include increased absorption, increased scattering and the decreasing feature size to match the lower wavelengths. Transmission spectra from two racetrack resonators with a bus to resonator gap, *g* = 250 nm were collected using a blue light source in order to evaluate the performance at wavelengths ~ 440 nm. Figure 4(a) shows the spectrum from a compact racetrack resonator (*R* = 5 um, *L* = 5 um). For the same sized resonator as those used in Figures 3 there is a decrease in FSR to 1.64 nm, resulting in an increased $n_g$ of 2.85. The increased bulk refractive index, *n* (see Figure S1) as well as the larger relative waveguide cross section at this wavelength account for this significant increase. For the resonant mode at 440.4 nm we see a $T_{min}$ = 0.56 and a *Q* = 3756 (Figure 4(b)). Based on the results in Figure 3, this device is operating in the undercoupled regime. From Equations 2 and 3 we can see that the resulting (*a*, *r*) values are 0.82 and 0.97 respectively. While *a* here is higher than those at green wavelengths, the *Q* factor for this particular mode is significantly higher than adjacent modes which are ~2800. Taking this into account we can recalculate round-trip amplitude transmission factor giving *a* = 0.77. From Equation 4 we establish a round trip power loss of 40% for this compact geometry. This is certainly expected due to the increased scattering for blue wavelengths.

Increasing the resonator radius to $R$ = 10 um results in the expected reduction in free spectral range (FSR), as shown in Figure 4(c). The quality factor extracted from the resonance at 440.9 nm in Figure 4(d) is relatively modest, with $Q$ = 2861. In general, increasing the resonator radius reduces bending losses and is therefore expected to increase the quality factor. However, in the present devices, the waveguide cross-section is relatively large compared to the operating wavelength, allowing higher-order TE and TM modes to propagate. We attribute the reduced $Q$ factor to overlapping resonances from these higher-order modes (Fig. 4(c)). The lower bending losses associated with the larger-radius resonator facilitate their propagation, whereas the stronger curvature of the smaller resonator shown in Figure 4(a) acts as a mode filter, preferentially suppressing higher-order modes[29]. Further optimisation of the waveguide geometry to enforce single-mode operation is therefore expected to significantly improve device performance at blue wavelengths.

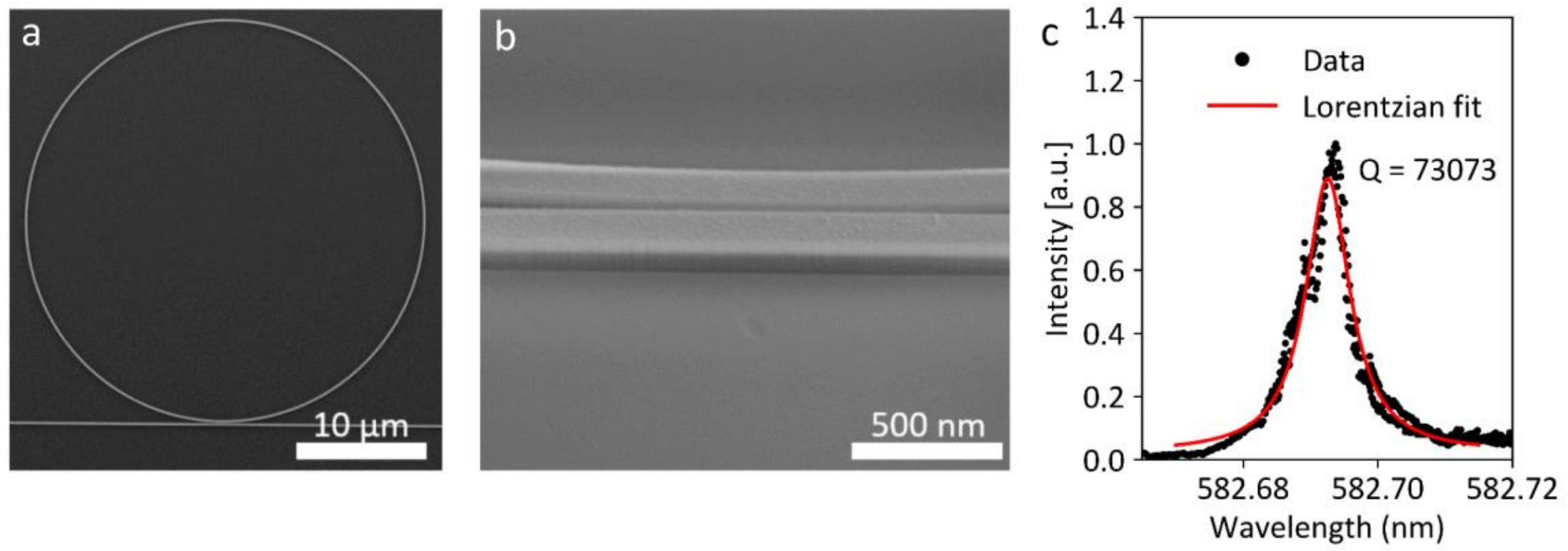

*Figure 5. (a,b) SEM images of a $Ta_2O_5$ ring resonator (R = 40 µm) evanescently coupled to a bus waveguide with a waveguide width of 220 nm, showing (a) the full device and (b) a magnified 52° tilted view of the coupling region from a similar device. (c) Scattered spectrum and Lorentzian fit of a resonant mode at 582.7 nm, from a larger radius ring resonator (R = 200 µm) yielding a quality factor of $Q > 7.3 \times 10^4$.*

To enhance device performance, larger ring resonators with radii of $R$ = 40 µm and 200 µm were fabricated to minimise bending losses and maximise the achievable quality factor. For the largest device, the coupling gap was increased to $g$ = 550 nm to intentionally operate in the undercoupled regime and maximise $Q$. An SEM image of a fabricated resonator ($R$ = 40 µm) is shown in Figure 5(a), while Figure 5(b) presents an angled close-up of the coupling region from a similar device, highlighting the high fabrication quality and minimal sidewall roughness.

Owing to the use of multimode excitation and collection fibres, together with the strongly undercoupled nature of the device, resonances were characterised through scattered-light measurements rather than the transmission measurements employed elsewhere in this work. A tunable dye laser was scanned across a resonance, and the collected scattering intensity was fitted with a Lorentzian profile, as shown in Figure 5(c). The extracted quality factor exceeds $7.3 \times 10^4$, corresponding to a resonance linewidth of $8.0 \times 10^{-3}$ nm. To the best of our knowledge, this represents one of the highest quality factors reported for $Ta_2O_5$-on-insulator resonators operating in the visible spectral range. The order-of-magnitude improvement in $Q$ relative to the devices presented in Figure 3 is attributed primarily to the substantial reduction in bending losses afforded by the larger resonator radius.

These results validate $Ta_2O_5$ as a highly capable material for visible photonics, offering the practical advantage of straightforward deposition and fabrication processes. The compact devices demonstrated robust performance across both blue and green wavelengths, despite being constrained by bending and scattering losses. As a demonstration of peak performance capabilities, the largest resonators achieved a quality factor exceeding $7.0 \times 10^4$ at green wavelengths, representing the highest reported *Q* factor for this material system in the visible spectrum. Furthermore, the versatility of the platform was demonstrated through the integration of hexagonal boron nitride (hBN) flakes, which were encapsulated within the structures using a two-step $Ta_2O_5$ deposition process to realise a hybrid material platform (details in Figures S4 and S5). Moving forward, geometric optimization of the waveguide and resonator dimensions represents a clear next step to maximize quality factors and minimize propagation losses, especially in the blue spectral region where efficient material platforms are limited. Such structural modifications would also provide an opportunity to improve the management of waveguide coupling behaviour across the visible spectrum. Ultimately, this platform offers a straightforward route towards high performance at challenging shorter wavelengths.

## Methods

### *Fabrication*

0.6 mm × 0.6 mm $Si/SiO_2$ (1 µm oxide) diced substrates were cleaned using sequential steps of ultrasonication in acetone and isopropyl alcohol and then dried under flowing nitrogen.

$Ta_2O_5$ was deposited with an AJA ATC-1800-E electron beam evaporator in both vacuum and oxygen environments. A4 Polymethymethacrylate (PMMA) was spun on the substrates as a positive electron beam resist. The patterns were written in with an Elionix ELS-F125 electron beam lithography system at 125 kV. The patterns were developed using a 1:3 mix of methyl isobutyl ketone (MIBK) to isopropyl alcohol (IPA). A Ti/Ni metal hard mask was then deposited with the AJA ATC-1800-E electron beam evaporator and lifted off in N-Methyl-Pyrrolidone (NMP) at 90°C. The $Ta_2O_5$ structures were etched by inductively coupled plasma reactive-ion etching (ICP-RIE) using a gas mixture of $SF_6$ /Ar. The Ti/Ni hard mask was removed with a piranha solution consisting of a 3:1 sulfuric acid and hydrogen peroxide. Electron microscope images were obtained using a scanning electron microscope (Thermo Fisher Scientific Helios G4).

### *Numerical Simulations*

3D FDTD simulations were undertaken using Ansys Lumerical software. The refractive index ($n$) of the $Ta_2O_5$ core was set to 2.17 ($k$ = 0) and 2.11 ($k$ = 0) for blue and green wavelengths, respectively, while $SiO_2$ was utilized as the substrate material. Perfectly matched layers (PML) were applied as boundary conditions around the simulation domain and total simulation time was $1.5 \times 10^4$ fs. In-built waveguide ports were used as injection and collection monitors for the $TE_{00}$ mode on the bus waveguide. Extracted Q factors were calculated from the decay envelope of the time-domain electromagnetic field.

***Spectral Transmission***

Transmission measurements were performed using a custom-built optical waveguide characterization setup. Thorlabs lensed fibers were mounted on three-axis (XYZ) micrometer stages to align each fiber with the grating couplers, at an angle of 10° from the vertical axis. Illumination was provided by LEDs centered at 440 nm and 555 nm. All spectra were collected using a fiber-coupled Andor Kymera 328i spectrometer equipped with an 1800 l/mm grating. For high-resolution scattering measurements, a tunable dye laser (Sirah Matisse 2 DS) was utilized as the excitation source and an Excelitas avalanche photodiode (APD) recorded photon counts.

**Acknowledgements**

The authors thank Rachel Grange for insightful discussions and valuable feedback on the experimental setup and results. We acknowledge financial support from the Australian Research Council (CE200100010, FT220100053, DP250100973), the Air Force Office of Scientific Research under award number FA2386-25-1-4044. The authors acknowledge the use of the fabrication facilities and scientific and technical assistance from the Sydney Nanoscience Hub Core Research Facility at the University of Sydney, being a part of the NCRIS-enabled Australian National Fabrication Facility (ANFF), and the UTS facilities, being a part of the ANFF-NSW node. We acknowledge Takashi Taniguchi (National Institute for Materials Science (NIMS) for providing the hBN crystals. Dr Jake Horder is gratefully acknowledged for fruitful discussions.